# Crawler for Image Acquisition from World Wide Web

R Rajkumar[1], Dr. M V Sudhamani[2]

[1]*Assistant Professor,* [2]*Professor and Head, Department of ISE, RNSIT*
*Bengaluru, Karnataka, India*

***Abstract:*** *Due to the advancement in computer communication and storage technologies, large amount of image data is available on World Wide Web (WWW). In order to locate a particular set of images the available search engines may be used with the help of keywords. Here, the filtering of unwanted data is not done. For the purpose of retrieving relevant images with appropriate keyword(s) an image crawler is designed and implemented. Here, keyword(s) are submitted as query and with the help of sender engine, images are downloaded along with metadata like URL, filename, file size, file access date and time etc.,. Later, with the help of URL, images already present in repository and newly downloaded are compared for uniqueness. Only unique URLs are in turn considered and stored in repository.*

*The images in the repository are used to build novel Content Based Image Retrieval (CBIR) system in future. This repository may be used for various purposes. This image crawler tool is useful in building image datasets which can be used by any CBIR system for training and testing purposes.*

## I. INTRODUCTION

In today's modern day world, the information is available in abundant. The Internet is the key to make it possible for such as rich information. The information are available in different forms over the World Wide Web (WWW). The WWW offers variety of web pages, in which the information stored in the form of texts, images, audios, and videos are available. To retrieve the right image from the large repository like WWW is difficult. In support of retrieval of right information, the search engines are used. The popular ones are Google, AltaVista, Bing, Yahoo, etc. These search engines use web crawlers that browse the entire WWW to collect the related information from corresponding URLs and store it in database. The web crawlers have enhanced the scope of searching for end users swiftly.

The content based image retrieval (CBIR) system is one of the domain in which an image need to be retrieved from a large datasets or WWW. Retrieval of the correct images is a real-time challenge for the CBIR system. A web crawler is a program that takes one or more seed URLs, downloads the linked web pages associated with such URLs [1]. The program may recursively continue to retrieve all the web pages identified by such links under the URL provided. The web crawler program search also based on a keyword too. The keyword can be associated with the image being retrieved as name of the file or caption or description of the file in the web page. These are found in the web pages links downloaded by the crawler. The crawler designed has to subsume some of the social responsibilities like ethics in crawling the websites frequently [15]. There are websites that warn for bots not to crawl into them and such server would implement the Robots Exclusion Protocol in a file called robots.txt [13]. Thus, the crawler needs to respect the protocol which is set by the crawler regulation standard adopted on web and allows minimal crawler ethicality, thereby avoiding network traffic to such servers.

## II. RELATED WORKS

The image search is of great interest and the techniques to do so is of great importance for the researchers across the globe. According to the survey [3] done, there are three categories of image search: (i) search by association (ii) aimed search and (iii) category search. The techniques provides excellent categories of image searching, but another important aspect to be noted by any search engine is that the freshness of the content or image [5]. The crawler implemented manages to get the image of interest but with intelligence that the same image is not retrieved and always is a new one. The majority of web pages across the Internet have become dynamic sites. Each time the crawler searches the dynamic pages, every time a new or fresh image is expected. The algorithms are written intelligently to know the difference between the current image to be retrieved and already existing image in the repository.

The duplication of retrieval of images by the web crawlers from the web pages is possible. There are standard techniques followed that provide [6, 8] filters like Hierarchical Bloom filters to remove the duplicate URLs before given as input for the crawler to parse and download the images. The WWW has enormous information associated with the query based keywords. The query keyword and the retrieved data or images may not be match 100% always. This is a limitation with a simple web crawler for images and thus need to add a meaningful keyword to be searched and is known as semantic search [7, 9]. The authors have proposed to extract metadata and meta-tags associated with the image retrieved from an URL and store it in the knowledge base (KB), so that the keyword to be searched can be filtered with the existing KB before crawling the web for faster and efficient searching.

The paper [4, 12] discusses about a special type of web crawler that migrates to the remote system called mobile crawler, where searching takes place and returns the relevant URLs. Such technique can be extended in parallel over the distributed systems, thereby reducing the network traffic and also computational power. But the limitation of such technique is that it might not have sufficient permission rights to parse the web links and send it from the remote machines.

### III. GENERAL ARCHITECTURE OF THE WEB CRAWLER

To extract the right image from the web is a tedious job. The ever growing Internet size has given horizons to develop powerful search engines. The web crawler eases the search for images on WWW. The crawler accepts the keyword as input scheduler and the parser in turn parses it to obtain the relevant URLs from the Internet and every URL is visited by it [10]. By identifying unique URLs, the crawler downloads the web pages. The general architecture of a web crawler is given below in fig 1:

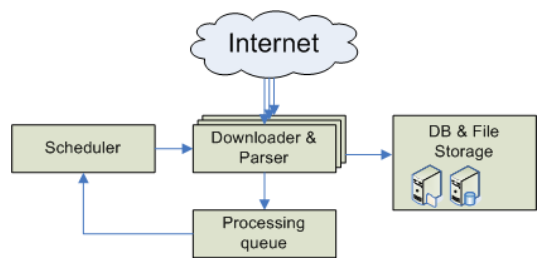

Fig. 1: General Architecture of Web Crawler

### IV. PROPOSED IMAGE CRAWLER ARCHITECTURE

The proposed image crawler architecture is implemented with a graphical user interface of web interface type; It consists of a keyword to be submitted by the user, the keyword is thus processed by the crawler module to search through any standard text based search engines like Google, Bing, Alta Vista or Yahoo. The URL results obtained from the search engines are parsed to check the valid URLs and its metadata. This source code information about the retrieved web pages is parsed by the crawler to download the images and its associated metadata. Relevant information like URL, file name, size, etc., is stored in MySQLdb in this work. Every time the crawler visits for the fresh image source, it checks for the redundancy in the database to ensure uniqueness of the URLs. Thus, redundancy and freshness of the web links are verified.

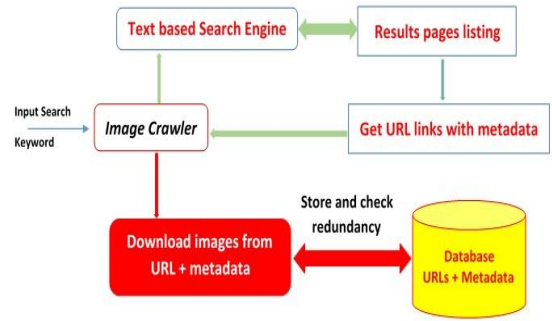

Fig. 2 Proposed Architecture of Image Crawler.

The above architecture diagram is implemented successfully to download the images to create a large dataset to support the content based image retrieval system. The work flow of the above architecture diagram is given as a pseudo code in the next section V.

### V. PSEUDO CODE OF THE WEB CRAWLER

The web crawler for image extraction is implemented as a client – server model. The GUI is deployed over the web server from which the crawler module is invoked. The algorithm and its processing are given as below:

```
Start the web server
Enter the keyword in the search text box
Web crawler module is invoked from the web server
The crawler module checks for Internet Connection
  If Internet connection successful then
    Processes the keyword to Text Search Engine
    Search Engine results are sent to parser
    Do until all the links are processed
      Checks the URL for valid image availability
      If Success in Retrieving Valid Image then
        Download the image and send the
        metadata + URL to the DB for storing
      Else if not Valid Image then
        Increment to the next URL
    Repeat the loop until all links are parsed
      successfully
  Else if Internet Connection fails then
    Display Internet Connection failure
Display the Search Window
End
```

The pseudo code written above shows the flow of the web crawler module. The pseudo code is written in python scripting language style-format. The tool checks for the network problems and display the appropriate message for processing the keyword search. The web crawler tool implemented downloads different images with as the purpose is to create a large dataset for training and testing the CBIR system under which the crawler is a component.

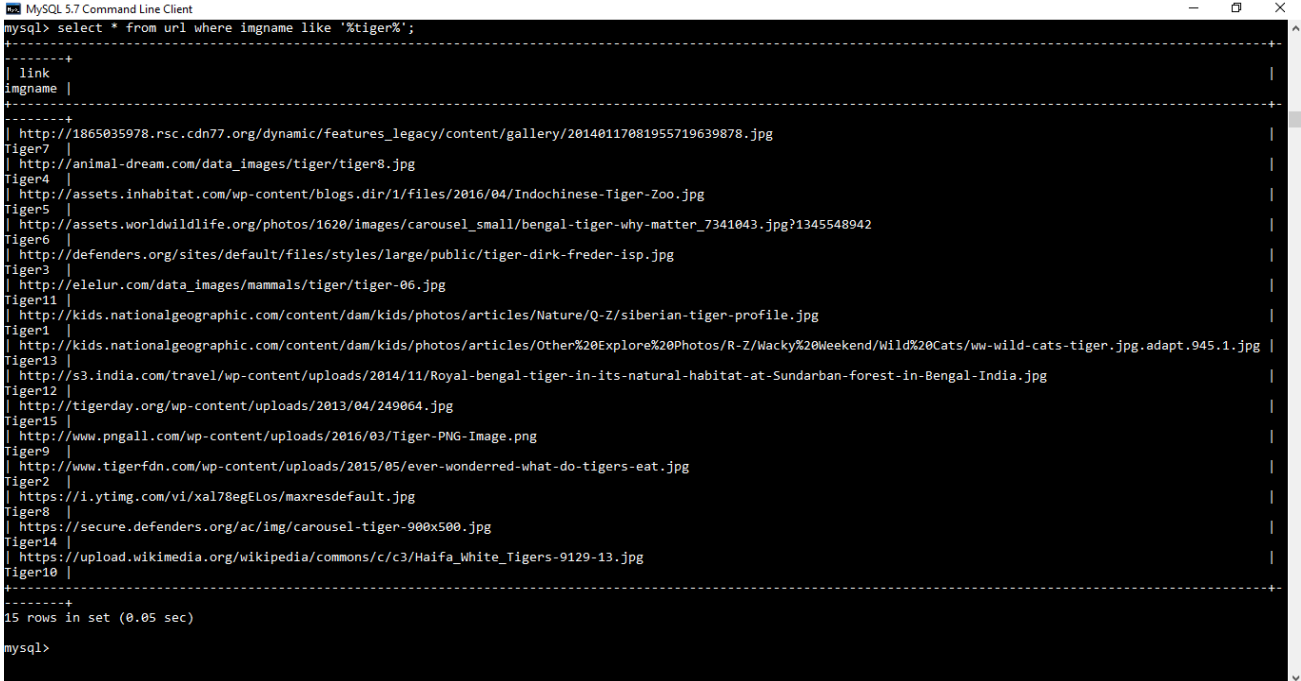

Fig 3. Snapshot of the downloaded image URLs and the file attributes.

## VI. EXPERIMENTAL RESULTS

The results obtained from the web crawler are stored in a database and a snapshot shown in fig. 3. This shows the URL from where the image is downloaded along file attributes. The web crawler tool is implemented using Python 2.7, with supporting libraries like selenium for parsing the web pages, Google custom search engine (CSE) API, beautiful soup for parsing the web pages for URL of the images and meta-data information extraction, *flask* is used as web server for client/server model and MySQLDB to store data. The images downloaded from the WWW are filtered with respect to the following standard image formats .jpg, .png, .bmp, .tiff and .gif. The following are the snapshots of results obtained during the retrieval process, fig.4 shows the first page in the web server to enter the query keyword of the image to be searched and downloaded. The fig. 5, fig. 6 and fig. 7 shows the result pages 1, 6 and 10 respectively obtained from WWW.

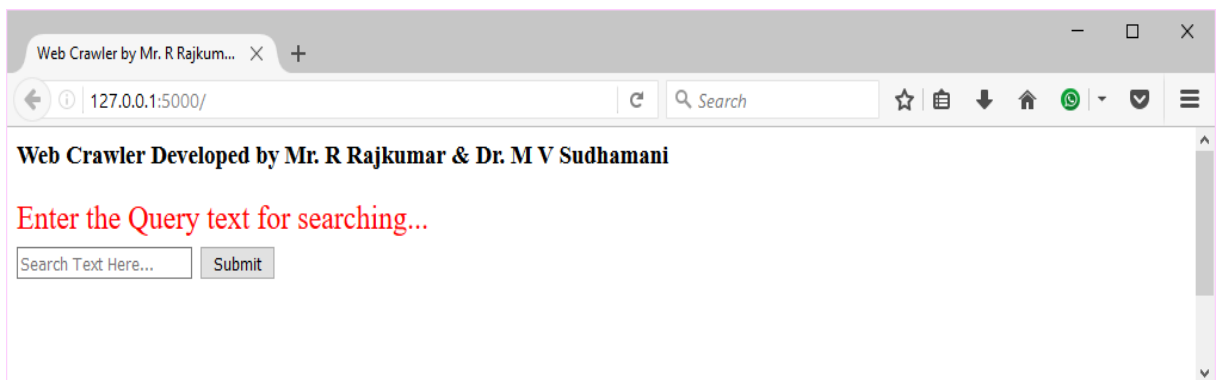

Fig. 4 Shows the GUI for search query

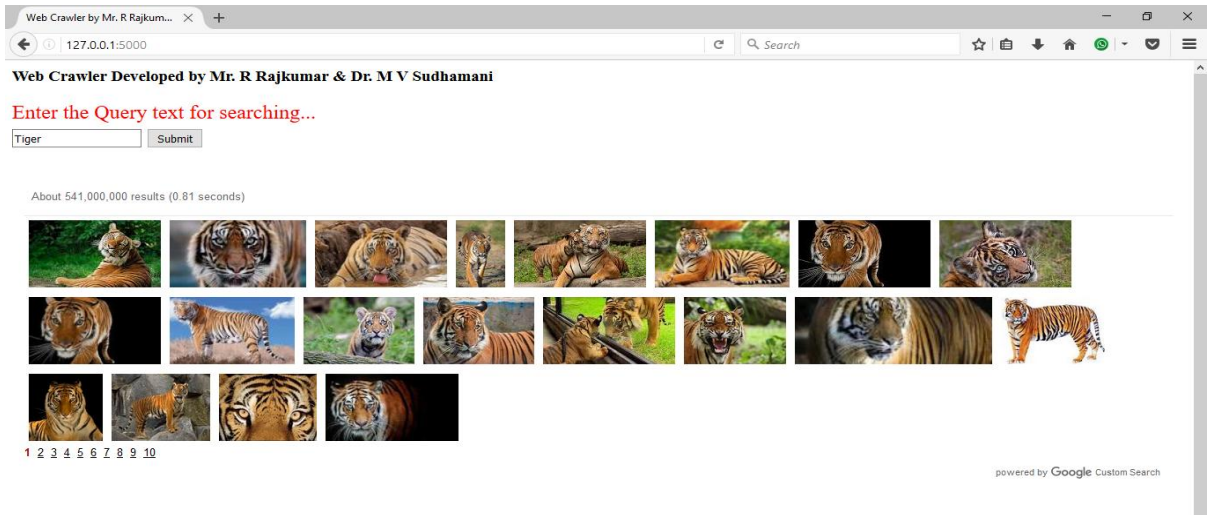

Fig. 5 shows the sample results for search query: Tiger page 1

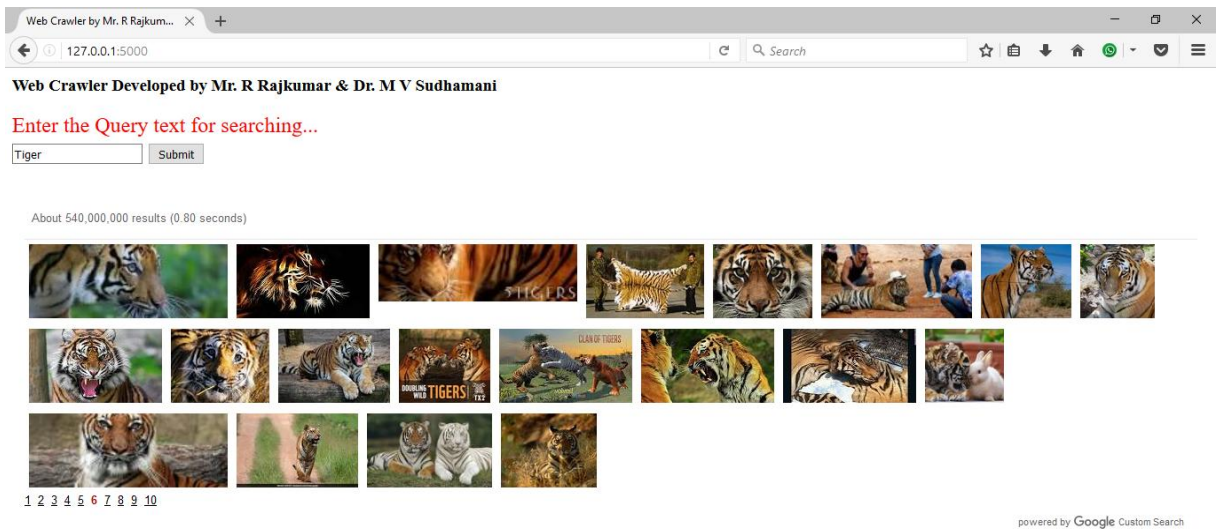

Fig. 6 shows the sample results for search query: Tiger page 6

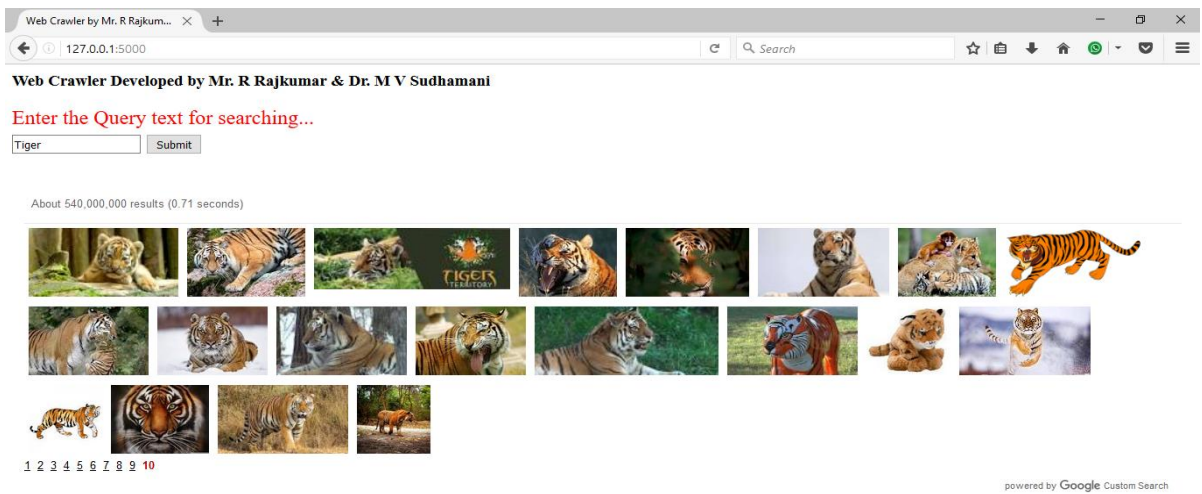

Fig. 7 shows the sample results for search query: Tiger page 10

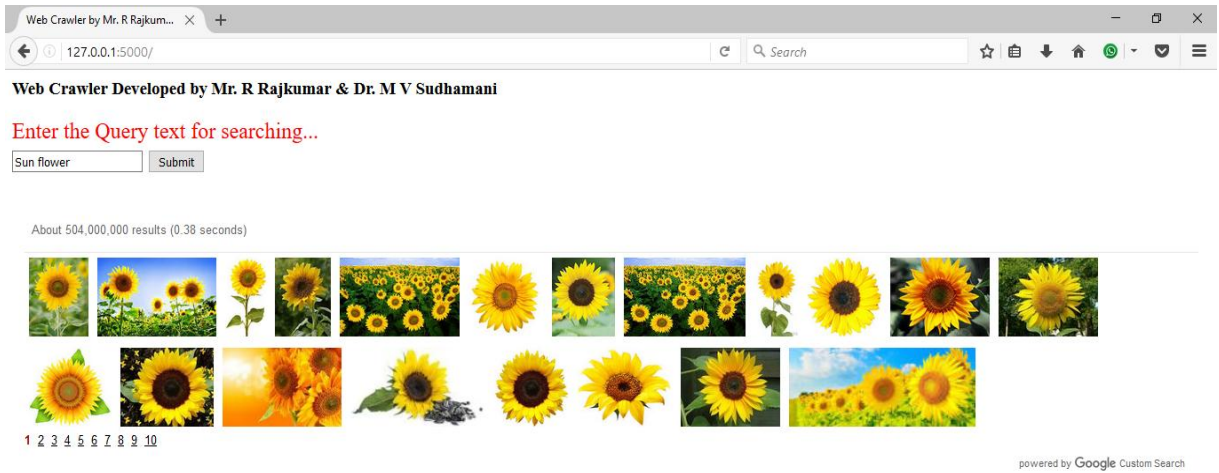

Fig. 8 shows the sample results for search query: Sun flower page 1

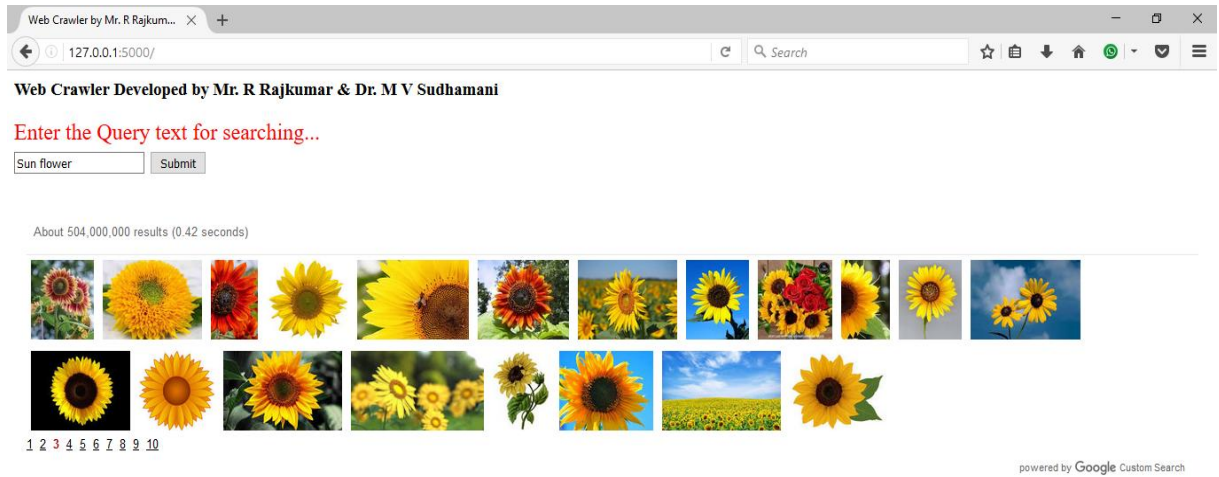

Fig. 9 shows the sample results for search query: Sun flower page 3

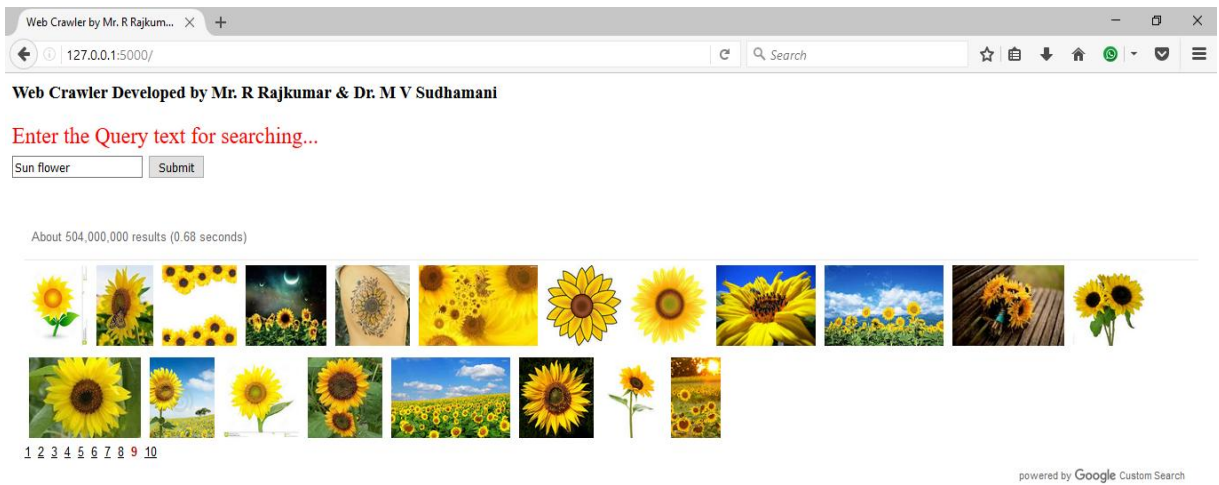

Fig. 10 shows the sample results for search query: Sun flower page 9

## VII. CONCLUSION

This paper presents an implementation of web crawler tool for image acquisition from WWW to create a repository of image datasets which are used in future to build CBIR systems. The experiments were carried out to test with different keywords. The crawler was tested for downloading more than 100 images for a given run. Likewise, a number of images were downloaded from different websites from the Internet. Almost relevant images are retrieved based on keyword query. In future work, in order to avoid spammed images from Ad-sites, the crawler has to be further refined with the help of appropriate filters. Here, an adaptive filtering technique can to be used for better performance of retrieval of relevant images by removal of unwanted sites or URLs being parsed by the crawler tool.

## REFERENCES

[1] Marc Najork, *Web Crawler Architecture*, Encyclopedia of Database Systems, Microsoft Research, Mountain View, CA, USA, 2009
[2] Bo Luo, Xiaogang Wang, and Xiaoou Tang, *A World Wide Web Based Image Search Engine Using Text and Image Content Features*, Internet Imaging IV, SPIE Vol. 5018, 2003.
[3] Ritendra Datta, Dhiraj Joshi, Jia Li, and James Z. Wang, *Image Retrieval: Ideas, Influences, and Trends of the New Age*, ACM Computing Surveys, pp. 5:1 – 5:60., Vol. 40, No. 2, April 2008.
[4] Abhinna Agarwal, Durgesh Singh, Anubhav Kedia, Akash Pandey, Vikas Goel, *Design of a Parallel Migrating Web Crawler*, IJARCSSE, Volume 2, Issue 4, ISSN: 2277 128X, April 2012.
[5] Junghoo Cho, Hector Garcia-Molina, *Effective Page Refresh Policies For Web Crawlers*, ACM Transactions on Database Systems, Vol. 28, No. 4, December 2003.
[6] Yingjun Wu, Han Huang, Xianzheng Zhou, Xiaobo Zhang, Feng Chen, *A Space-saving URL Duplication Removal Method for Web Crawler*, Journal of Information & Computational Science, http://www.joics.com, May 2012
[7] Anthoniraj Amalanathan and Senthilnathan Muthukumaravel, *Semantic Web Crawler Based on Lexical Database*, IOSR Journal of Engineering, ISSN: 2250-3021, Vol. 2(4) pp: 819-823, April 2012.
[8] Peng Yang, Hui Li, Qingshan Liu, Lin Zhong, Dimitris Metaxas, *Content Quality Based Image Retrieval With Multiple Instance Boost Ranking*, MM'11, November 28–December 1, 2011, Scottsdale, Arizona, USA, ACM 978-1-4503-0616-4/11/11, 2011.
[9] Chetna, Harpal Tanwar, Navdeep Bohra, *An Approach to Reduce Web Crawler Traffic Using ASP.NET*, International Journal of Soft Computing and Engineering (IJSCE) ISSN: 2231-2307, Volume-2, Issue-3, July 2012.
[10] Swati Ringe, Nevin Francis, Palanawala Altaf, *Ontology Based Web Crawler*, International Journal of Computer Applications in Engineering Sciences, ISSN: 2231-4946, VOL II, ISSUE III, Sept 2012.
[11] Mini Singh Ahuja, Dr. Jatinder Singh Bal, Varnica, *Web Crawler: Extracting the Web Data*, International Journal of Computer Trends and Technology (IJCTT), ISSN: 2231-2803, Vol: 13 No: 3, Jul 2014.
[12] Niraj Singhal, Ashutosh Dixit, R. P. Agarwal, A. K. Sharma, *Regulating Frequency of a Migrating Web Crawler based on Users Interest*, International Journal of Engineering and Technology (IJET), ISSN : 0975-4024 Vol: 4 No: 4, Aug-Sep 2012.
[13] Ram Kumar Rana, Nidhi Tyagi, *A Novel Architecture of Ontology-based Semantic Web Crawler*, International Journal of Computer Applications (0975 – 8887), Volume 44– No.18, April 2012.
[14] Purohit Shrinivasacharya, Dr. M V Sudhamani, *An Image Crawler For Content Based Image Retrieval System*, International Journal of Research in Engineering and Technology (IJRET), eISSN: 2319-1163 | pISSN: 2321-7308, Volume: 02, Issue: 11, Nov - 2013.
[15] Sun. Y, Council G. Isaac and Giles C. Lee, *The Ethicality of Web Crawlers*, in the proceedings of 2010 IEEE/WIC/ACM International Conference on Web Intelligence and Intelligent Agent Technology,.(pp: 668-675), Toronto Canada, August 2010.
[16] Ms. Sayali.S.Pawar, Prof. R.S.Chaure, *A New Trend Content-Based Image Retrieval Technique used in Real Time Application*, International Journal of Advanced Research in Computer Science and Software Engineering, ISSN: 2277 128X, Volume 4, Issue 6, June 2014.